\documentclass[namedreferences]{SolarPhysics}
\usepackage[optionalrh]{spr-sola-addons} % For Solar Physics
\usepackage{graphicx}        % For eps figures, newer & more powerfull
\usepackage{color}           % For color text: \color command
\usepackage{url}             % For breaking URLs easily trough lines
\newcommand{\etal}{{\it et al.},~}
\newcommand{\etalb}{{\it et al.}~}
\newcommand\ie{i.e.,~}
\newcommand\RS{$\log (R/S)$}
\renewcommand\S{$\log (w)$~}
\renewcommand\H{Hurst exponent}
\newcommand\T{time series}

\newcommand\ro{ R\"ossler}
\newcommand\lo{Lorenz}
\newcommand\eqn{\end{equation}\noindent}
\newcommand\eqnr{\end{eqnarray}\noindent}
\newcommand\beqr{\begin{eqnarray}}
\newcommand\beq{\begin{equation}}

\newcommand{\aap}{    {\it Astron. Astrophys.}}

\newcommand{\prl}{    {\it Phys. Rev. Lett.}}
\newcommand{\pra}    {{\it Phys. Rev. A }}

\begin{document}

\begin{article}

\begin{opening}

\title{Nonlinear Time Series Analysis of Sunspot Data}

\author{Vinita~\surname{Suyal}$^{1}$\sep
        Awadhesh~\surname{Prasad}$^{1}$\sep
        Harinder~P.~\surname{Singh}$^{1}$
       }
\runningauthor{Suyal et al.}
\runningtitle{Nonlinear Time Series Analysis of Sunspot Data}

   \institute{$^{1}$ {Department  of Physics  and Astrophysics,
University of Delhi, Delhi 110007, India\\
email:hpsingh@physics.du.ac.in}}

\begin{abstract}
This paper deals with the analysis of sunspot number time series
using  the \H. We use the rescaled range ($R/S$) analysis to estimate the
\H~for 259-year and $11\,360$-year sunspot data.
The results show a varying degree of  persistence over shorter
and longer time scales corresponding to distinct values of the \H.
We explain the presence of these multiple \H s~by their resemblance
to the deterministic chaotic attractors  having  multiple centers
of rotation.

\end{abstract}
\end{opening}
%-------------------------------------------------

\section{Introduction}
     \label{S-Introduction}

Strong magnetic field present in the sun's outer regions is
manifested by complex temporal dynamics, {\it e.g.}, sunspots, solar
wind velocity and solar flares. Magnetic activity manifests itself
most clearly in sunspots. It has been found that chromospheric
flares show a very close statistical relationship with sunspots
\cite{bray}. Long term variations of solar activity may cause
climatic changes on earth, whereas short term variations may be
accompanied by fluctuations of certain
 meteorological parameters \cite{wittmann}.
Besides its 11-year fundamental periodicity, solar activity
as measured by relative sunspot number shows quasiperiodic variations
with period ranging from 2 to 1100 years \cite{michelson,kimura,turner,
kiral,zhukov,cole}.

Alongside several periodicities, the solar activity also exhibits
irregular fluctuations and these fluctuations were first assumed to
be determined by the short term variation with a random distribution
\cite{ruzmaikin}. The rediscovery of the  grand minima of solar
activity \cite{eddy} led to a re-examination of the nature
 of the non-periodic part of the variations of the sun's activity.
Solar activity in the frequency range from 100 to 3000
years includes an
important continuum component in addition the well-known
periodic variations \cite{ruzmaikin}.

The \H ~is a parameter that quantifies the persistent or
anti-persistent (past trends tend to reverse in future) behavior of
a time series. It determines whether the given time series is
completely random or has some long-term memory. Ruzmaikin
\etalb(1994) examined whether or not the nonperiodic variations in
solar activity are caused by a white-noise, random process. They
evaluated the \H ~for a \T ~of $^{14}$C data from 6000 BC to 1950
AD. They find a \H ~of $\approx 0.8$ indicating a high degree of
persistence in the variations of solar activity. Kilcik \etalb
(2009) used the monthly ISSN (international sunspot numbers) data
for the last 3000 years to evaluate the \H ~with a view to predict
the sunspot activity for solar cycle 24. Xapsos \etalb(2009) used
the reconstructed sunspot numbers for the past $11\,360$ years by
Solanki \etalb(2004) to find the \H ~of $\approx 0.8$ and also
showed the evidence of 6000-year periodicity in the reconstructed
sunspot numbers. In all the above studies involving the  Hurst
analysis to understand the persistent behavior of the sunspot data,
a single \H ~was estimated although there are many scaling regimes
which give different \H s. In this paper, we use the Hurst analysis
on 259-year and $11\,360$-year data sets and find multiple \H s in
each \T. We explain the presence of multiple \H s for a single \T~
using systems from deterministic chaotic dynamics with multiple
centers of rotation. Our results conclude that estimating a single
\H ~from the data where different linear scaling regimes exist may
be improper.

 In Section~\ref{S-rs} we review the $R/S$ method to calculate
 the Hurst exponent of a time series. The results for sunspot data are
 discussed in Section~\ref{S-resultastro}. Results for different
 chaotic models having one or more centers of rotation in phase space
 are given in Section~\ref{S-resultmodels}. The conclusions are given
 in Section~\ref{S-conc}.

\section{Hurst Analysis and $R/S$ Measure} %%%%%%%%%%%%%%%%%%%%%%%%%%%%%%%%%%%%%%%%
      \label{S-rs}

The $R/S$ method to find the \H~was proposed by Mandelbrot
 and Wallis (\citeyear{mandelbrot}b) which
 can be summarized as follows:

 Let $X_{i},~ i=1,2,...,N$, be an observed time series whose \H~is to be computed.
 Let us now choose a parameter $w$ (temporal window such that $w_t \le {w} \le {N}$ where $w_t$ is
the Theiler window \cite{thw})
 and consider the subsets of the data
$x_{i},~i=t_{0}, t_{0}+1, t_{0}+2,...,
t_{0}+w-1$, where $1 \le t_{0} < {N-w+1}.$

We then denote the average of these subsets as

$$\bar{x}(t_{0},w) = \frac{1}{w}\sum_{i=t_0}^{t_0+w-1}{x_i}.$$

Let $S(t_0,w)$ be the standard deviation of $x_i$, during the window
$w$ {\it \ie}

$$ S(t_0,w)=\left[ \frac{1}{w-1}\sum_{i=t_0}^{t_0+w-1}\left\lbrace
x_i -\bar{x}(t_0,w)\right\rbrace ^2\right] ^{\frac{1}{2}}.$$

Next, new variables $y_i,~ i=1,2,...,w, $ and \textit{range} $R$
are defined as

$$y_i(t_0,w) = \sum_{k=t_0}^{t_0+i-1}[x_k-\bar{x}(t_0,w)],$$

$$R(t_0,w) = \max_{1\leqslant i \leqslant w}y_i(t_0,w) - \min_{1\leqslant i
 \leqslant w}y_i(t_0,w),$$
\noindent which allows one to define \textit{Rescaled range} measure $R/S$ as

$$(R/S) (t_0,w) = \frac{R(t_0,w)}{S(t_0,w)}.$$

Taking $ t_0 = 1, 2, ..., N-w+1$, and computing $(R/S)(t_0,w)$ for
time lag $w$ the rescaled range for the time lag $w$ is finally
written as the
 average of those values
\begin{equation}
R/S = \frac{1}{N-w+1} \sum_{t_0} (R/S)(t_0,w).
\end{equation}

It has been observed that the rescaled range $(R/S)$ over a time
window of width $w$ varies as a power law:
\begin{equation}
(R/S)_w = k\,w^H, \nonumber
\end{equation}

\noindent where $k$ is a constant and $H$ is the Hurst exponent. To estimate
the value of the Hurst exponent, $R/S$ is plotted against $w$  on
log-log axes. The slope of the linear regression gives the value of
the Hurst exponent. If the time series is purely random then the
\H~$(H)$ comes out to be $0.5$. If $H > 0.5$, the \T ~covers more
`distance' than a random walk, and is a case of persistent motion,
while if $H < 0.5$, the \T ~covers less `distance' than a random
walk it shows the anti-persistent behavior in \T. However, for
periodic motions, the \H ~is $1$.

\section{Sunspot Data and $R/S$ Analysis} %%%%%%%%%%%%%%%%%%%%%%%%%%%%%%%%%%%%%%%%
      \label{S-resultastro}

\begin{figure}
\centerline{\includegraphics[width=1.0\textwidth,clip=]{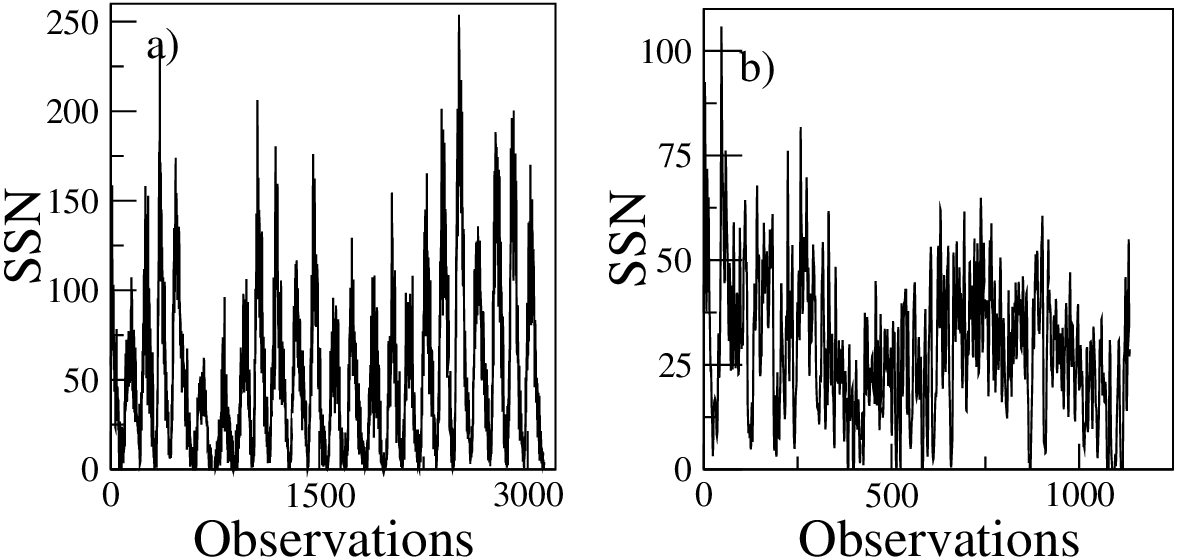}
}\caption{ a) Monthly sunspot record from January 1749 to December 2008.
           b) Reconstructed sunspot record from Solanki \etalb(2004).
The time period covers $11\,360$ years with observations in 10-year increments.}
\label{F-spots}
\end{figure}

\begin{figure}
\centerline{\includegraphics[width=1.0\textwidth,clip=]{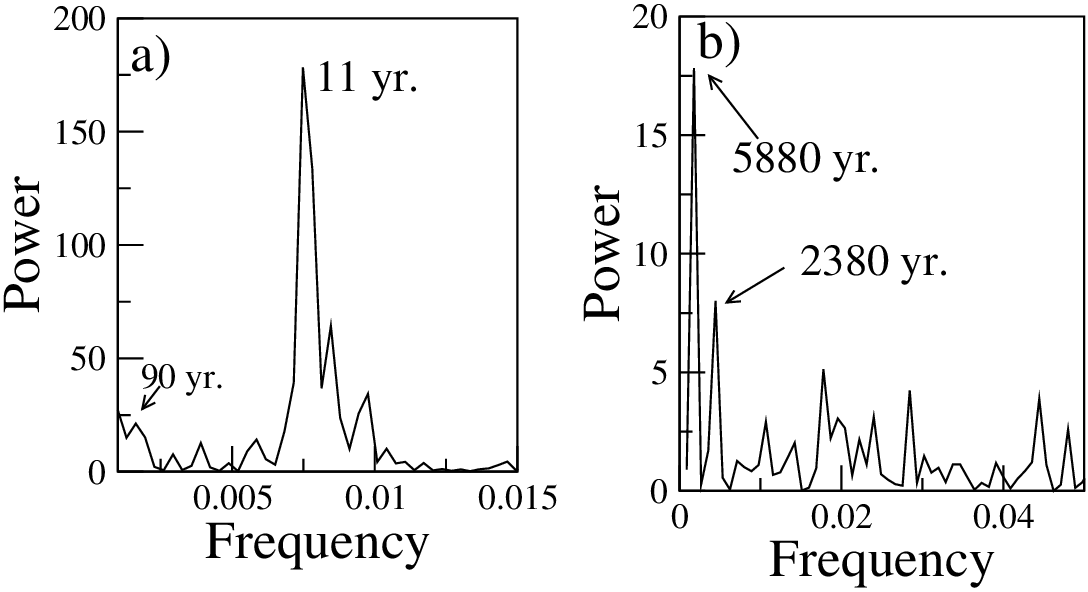}
}\caption{ Power spectra for a) monthly sunspot
record from January 1749 to December 2008.
           b) Reconstructed sunspot record from Solanki \etalb (2004).
The time period covers $11\,360$ years with observations in 10-year increments.}
\label{F-spower}
\end{figure}

\begin{figure}
\centerline{\includegraphics[width=0.5\textwidth,clip=]{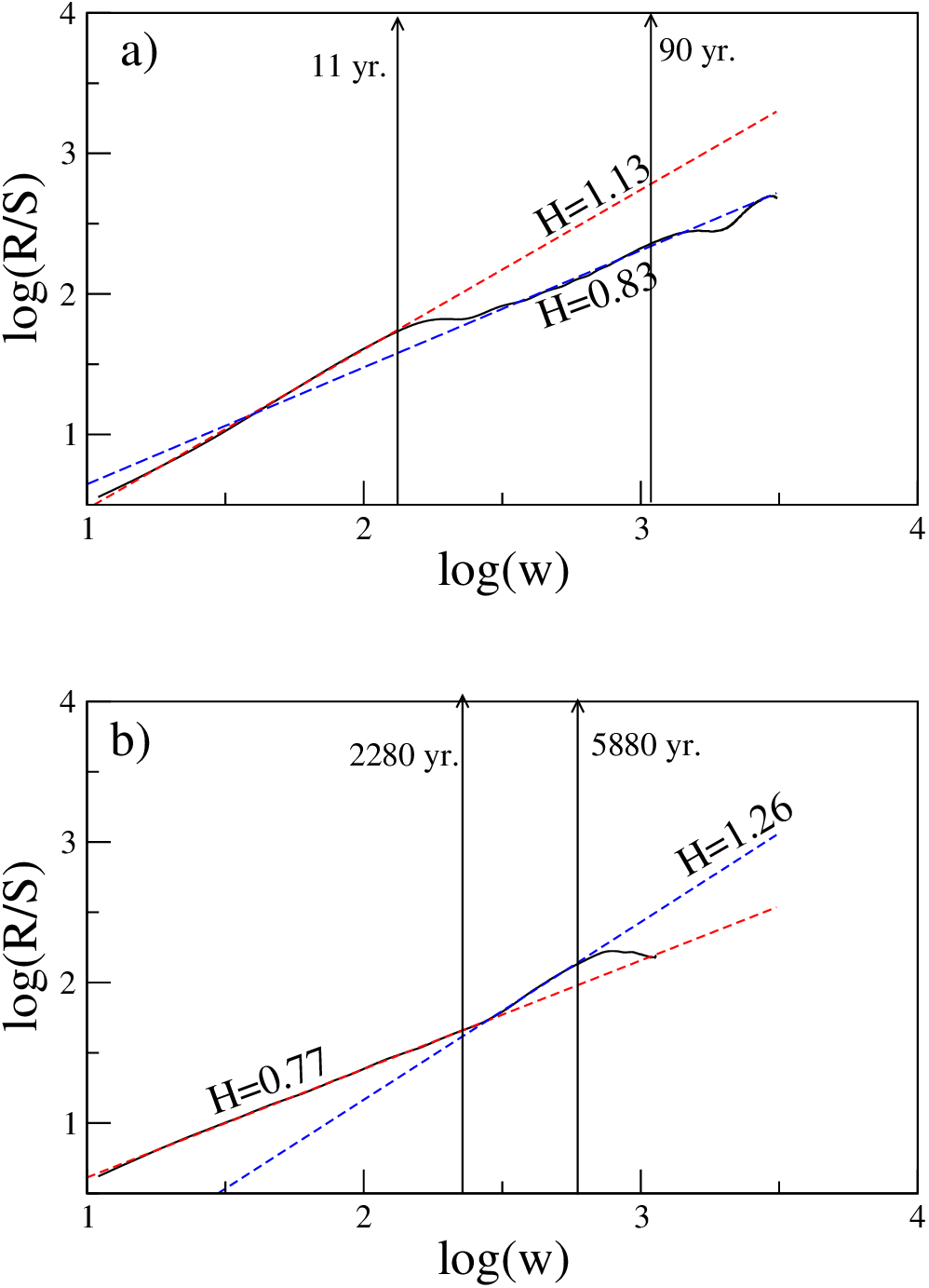}
}\caption{{ \RS} vs.~{\S} for sunspot number a) from
January 1749 to December 2008 b) reconstructed sunspot record of
$11\,360$ years. The Hurst exponents with corresponding standard
deviations are a) 1.13$\pm$0.003; 0.83$\pm$0.003 b) 0.77$\pm$0.001;
1.26$\pm$0.004.} \label{F-shurst}
\end{figure}

Sunspot number (SSN)
is continuously changing in time and
constitutes a time series. Figure~\ref{F-spots}(a) shows the
monthly-averaged sunspot numbers from
Sunspot Index Data Center SIDC (http://www.sidc.be/sunspot-data)
from 1749 up to the present. The power spectrum of this data set
is shown in Figure~\ref{F-spower}(a).
We further estimate the \H ~for this \T ~using the $R/S$ method.
Figure~\ref{F-shurst}(a) shows the log-log plot for the 259-year data
showing the presence of  multiple Hurst exponents
for this time series.
The first bending is due to the prominent 11-year cycle
and  the estimated \H~1.13 in the linear regime before this bending.
The next prominent bending is at roughly 90 years showing
the Wolf-Gleissberg cycle.
The estimated \H ~in the second linear regime
between 11 years and 90 years is 0.83.


We further analyze the reconstructed sunspot \T ~of $11\,360$ years
(Solanki \etal 2004) shown in Figure~\ref{F-spots}(b).
Figure ~\ref{F-spower}(b) shows the power spectrum
of this \T ~indicating the presence of two long-term periodicities
at nearly 2300 and 6000 years. The $R/S$ analysis
also shows two bendings around these periods (Figure ~\ref{F-shurst}(b)).
 Again, for this data set we get two linear regimes,
 the \H ~of the first part being 0.77 and the second part being
1.26.

If we were to calculate a single \H ~for these two \T ~we will
get $H \approx 0.9$ for 259-year and $H \approx 0.8$ for $11\,360$-year data
(Figure \ref{F-shurst}(a) and (b) respectively). These values of the \H~
agree with the previously estimated values of {\it H} in the literature \cite{ruzmaikin,
alana,xapsos}.
However, as we have demonstrated, the log-log plots of sunspot data
show two remarkably distinct scaling regimes
and hence estimating only one \H ~may be improper. In the next section
we examine in  detail the reasons for
these different scaling regimes by using examples of a variety
of \T~of standard chaotic systems.

\section{Chaotic Models and $R/S$ Analysis} %%%%%%%%%%%%%%%%%%%%%%%%%%%%%%%%%%%%%%%%
      \label{S-resultmodels}

\begin{figure}
\centerline{\includegraphics[width=0.8\textwidth,clip=]{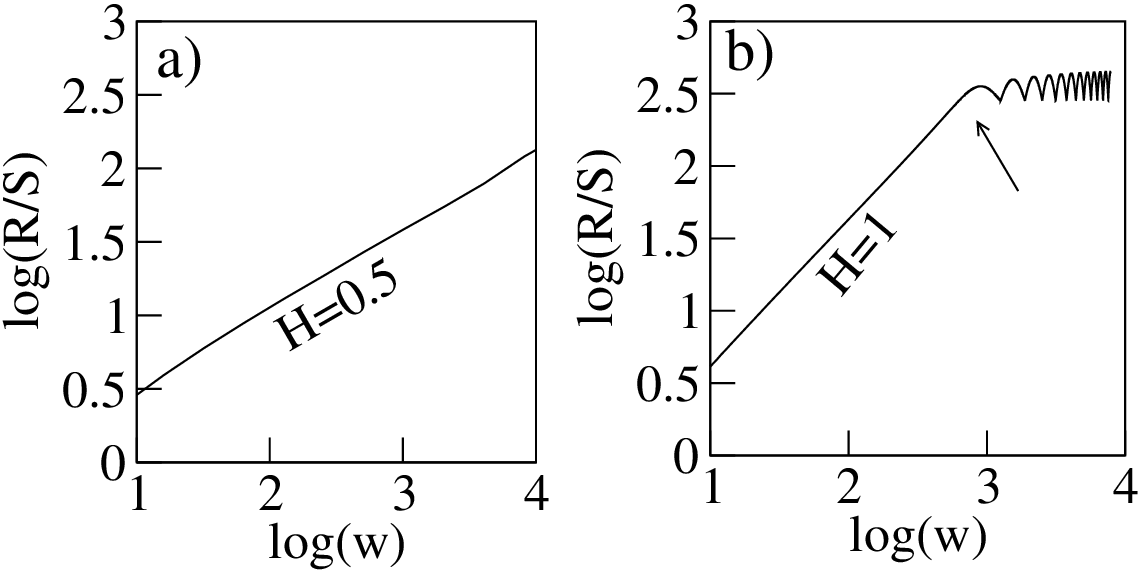}
}\caption{ Plot of  \RS ~vs.~\S for a) random and b)
periodic time series.}
 \label{F-extreme}
\end{figure}

In order to understand the behavior of different dynamical
systems using the
\H, we take two extreme examples of time series generated from
random and periodic motion respectively. First we take a random time
series with $100\,000$ data points distributed uniformly in the range
$[0,1]$. Using the $R/S$ method we compute the Hurst exponent as
discussed in Section~2 and
 find the \H~$H = 0.5$ as given by the slope of the
 line in Figure~\ref{F-extreme}(a), which  is expected
for a purely random \T~\cite{mandelbrot2}.

We further consider  the other extreme case of a purely periodic
signal of period one generated from
$\sin (x)$, with $x$ in the range $[0,1000]$ with step size $0.01$.
 Figure~\ref{F-extreme}(b) shows the plot
 of \RS~vs.~\S and the slope of the linear regime in this case is $1$.
One prominent difference between these two extreme cases of a purely
random signal and a periodic signal (Figure \ref{F-extreme}(a) and
(b) respectively), is that for a purely random signal, the plot of \RS~vs.
\S has a constant slope, while for a periodic signal it gets
saturated and starts oscillating after a certain value of $w$ (marked
by an arrow). The value where the bending begins corresponds to the
period of oscillation $T=2\pi/0.01$, and has been verified by the
power spectrum. The bending, therefore, gives us
information about the
frequency of the given signal.

\subsection{Chaotic Time Series}

\begin{figure}
\centerline{\includegraphics[width=0.7\textwidth,clip=]{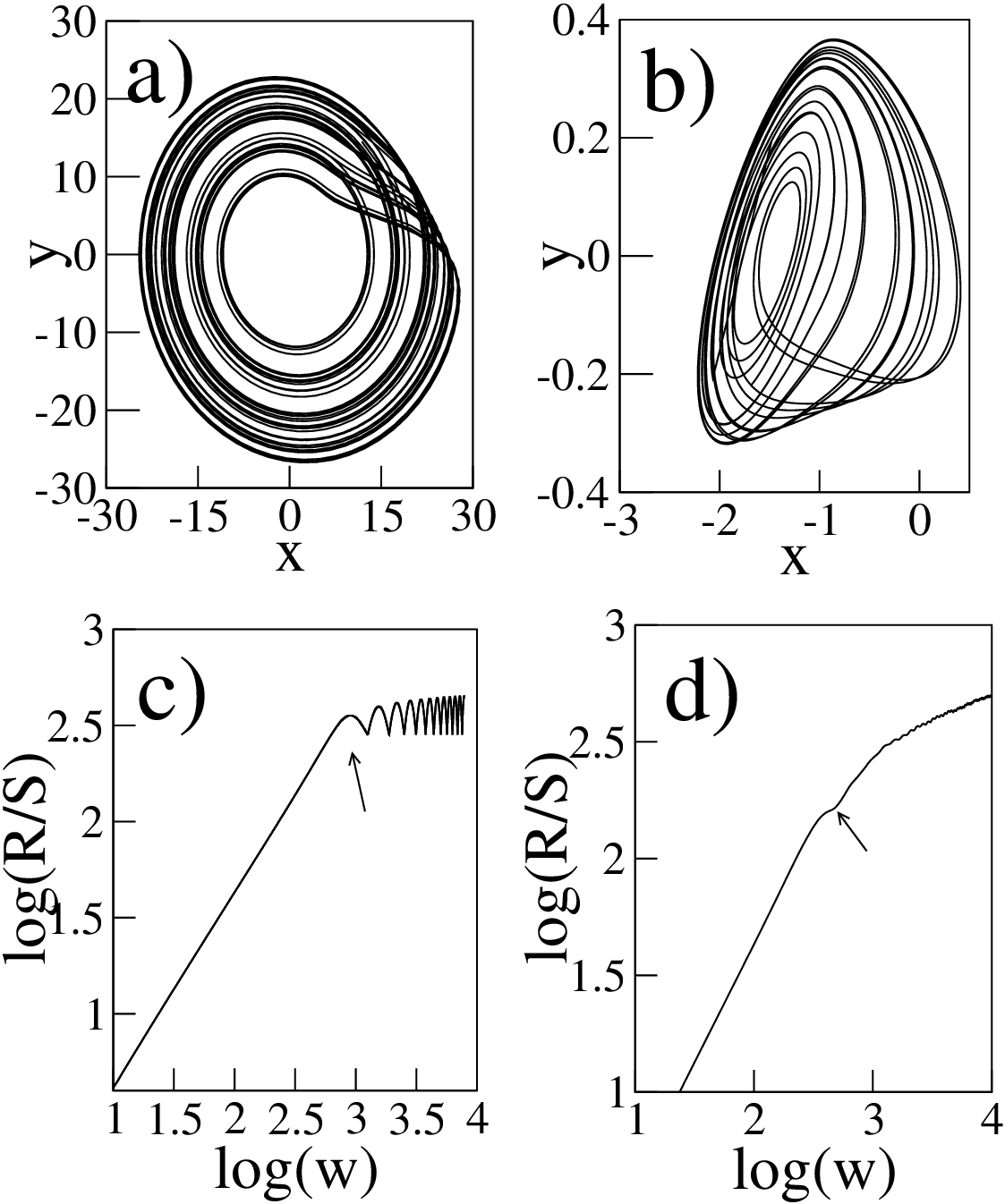}
}\caption{Plot of trajectories  in the $x-y$ plane  for chaotic motion
of (a) \ro, b) Chua oscillators having  single center of rotation.
Corresponding plots of \RS~vs.~\S are given in panels c) and d)
respectively. } \label{F-single}
\end{figure}

In the past few decades, chaotic motion, observed in deterministic
systems, has received  great attention due to its presence in
systems from physical, chemical, biological, ecological,
physiological to social sciences. These chaotic motions are
temporally aperiodic and are strange because they have a fractal
geometry.  Since chaotic motion is neither periodic nor random we
can expect its behavior to be in-between two extreme cases of
randomness and periodicity. Therefore, we can also
expect the \H ~to be
different from $0.5$ and $1$.
 An example of a system exhibiting chaotic motion is the
celebrated  \ro ~attractor \cite{rossler},
\begin{eqnarray}
\dot x & =& -y-z ,\nonumber\\
 \dot y &=& x+ay ,\nonumber\\
 \dot z& =& b+z(x-c),
\end{eqnarray}
\noindent
where  $a$, $b$ and  $c$ are control parameters. A chaotic trajectory
 (with $a=0.1$, $b=0.1$~and $c=18.0$)
 in the $x-y$ plane is shown in
 Figure~\ref{F-single}(a) which rotates around an unstable
period-one fixed point
$((c - d )/2,~ (-c + d)/{2a},~ (c - d)/{2a})$ where $d = \sqrt{c^{2}
-4ab}$~. The plot of~\RS ~vs.~\S (Figure \ref{F-single}(c))
 shows that there is linear regime (before marked
arrow) after which it gets saturated.
The slope of linear region gives
$H \approx 1$. The first bending of the above-mentioned curve gives the
frequency which matches with the frequency obtained
from power spectrum or peak
to peak analysis of the amplitude. This
system is dissipative and
has to be bounded in the sub-phase
space~\cite{rossler} and the
bending here corresponds to the
folding nature of the dynamics.

In order to see if the behavior is
replicated in another similar
system, we consider the Chua oscillator (Chua~{\it et al.}, 1993),
\begin{eqnarray}
\nonumber
\dot x &=& c_1(y-x-p(x)),\\
\nonumber
\dot y &=& c_2(x-y+z),\\
\dot z &=& -c_3y,
\label{eq:chua}
\end{eqnarray}

\noindent where $p(x)$ is defined as,
$$p(x)=m_1x+\left((m_0-m_1)(\mid x+1\mid - \mid x-1 \mid)\right)/2.$$
We take $c_1 = 15.6$, $c_2 = 1, m_0=-8/7,~m_1 = 5/7$ and first select
the parameter $c_3=33$ such that the motion will be a single-scroll type
as shown in Figure~\ref{F-single}(b) (there is a symmetric attractor
also, depending upon the initial conditions). The \RS~vs.~\S plot
for variable $x$ is shown in Figure \ref{F-single}(d) and the \H ~is
found to be $H \approx 1$. This also shows saturation at the average time
period (shown by arrow). These two examples of chaotic dynamics,
\ro ~and Chua systems,  clearly demonstrate that whenever there is a
single center of  rotation the \RS~vs.~\S plot shows a linear
scale only up to the average period of the attractor after which it
saturates.

In nature one may come across many dynamical systems which are
chaotic and their trajectories rotate around more than one center of
rotation. In order to see the behavior of  \RS~vs.~\S plot, we
first consider the Lorenz system \cite{lorenz},
\begin{eqnarray}
\dot x &=& \sigma(y-x),\nonumber\\
 \dot y &=& x(\rho-z)-y,\nonumber\\
 \dot z &=& xy-\beta z.
\label{eq:lor}
\end{eqnarray}

\begin{figure}
\centerline{\includegraphics[width=0.7\textwidth,clip=]{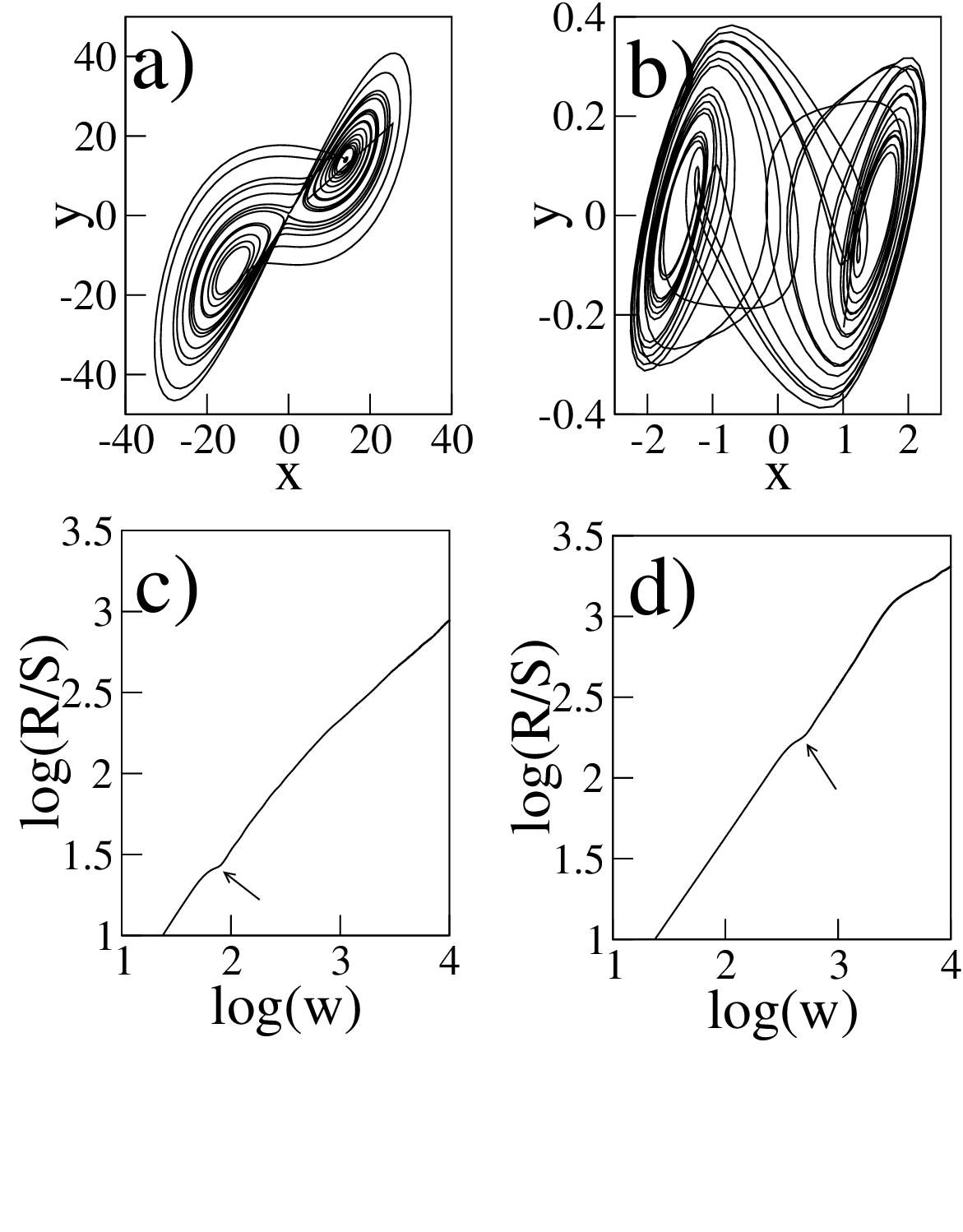}
}\caption{Plot of  trajectories  in the $x-y$ plane  for chaotic motion
of a)
 \lo ~and b) Chua oscillators having  multiple centers of rotation.
 Corresponding plots of \RS~vs.~\S are given in panels c) and d) respectively.}
\label{F-double}
\end{figure}

\noindent
Shown in Figure~\ref{F-double}(a) is the chaotic
trajectory in the $x-y$ plane
at parameter values $\sigma=16$, $\rho=50$ and $\beta=4$.
This clearly shows that the trajectory rotates
for some time around the fixed points:\\

\noindent $( \sqrt{\beta \left( \rho -1\right)} ,
\sqrt{\beta \left( \rho -1\right)},\left( \rho -1\right)) $
 and $( -\sqrt{\beta \left( \rho -1\right)} ,-\sqrt{\beta \left
( \rho -1\right)},\left( \rho -1\right) )$,\\

\noindent while some time it
rotates around all of these fixed points
(including $(0,0,0)$).
The \RS~vs.~\S plot in
Figure~\ref{F-double}(c) shows that there are two regimes of linear
scaling. The slope of the first part of the linear region gives
$H=0.93$ and the second part gives $H=0.64$. The first linear regime
corresponds to the trajectory rotating about individual fixed points
while the second is for all three fixed points. This indicates that
the dynamics around individual fixed points is different from that of
the combined one. Therefore taking the slope of individual regimes can
give more details of the intrinsic dynamics. This method also
allows us to estimate intrinsic frequencies of the system.

In order to see this behavior in another system where chaotic motion
contains many unstable fixed points around which the trajectory
revolves, we consider the Chua circuit represented by
Equation~(\ref{eq:chua}). Figure~\ref{F-double}(b) shows the
trajectory having a double-scroll chaotic motion (Chua system at
$c_3=28$). Similar to the \lo ~system, this system confirms the
existence of two different regimes of linear scaling having the \H
s, $H=1$ and $1.1$. These two examples  confirm that for multiple
centers of rotation, there are many regimes of linear scaling for
which distinct \H s can be estimated.

\section{Conclusions} %%%%%%%%%%%%%%%%%%%%%%%%%%%%%%%%%%%%%%%%
      \label{S-conc}

In this paper, we use the Hurst analysis on 259-year and
$11\,360$ -year data sets and find multiple \H s in each \T.
We explain the presence of multiple \H s in a
single \T ~using systems from deterministic chaotic dynamics
with a single center of rotation
(\ro~and single-scroll Chua oscillators) as well as
multiple centers of rotation (\lo~and double-scroll Chua oscillators).
We have shown that in the sunspot data, two distinct
linear scaling regimes exist for which two distinct \H s
could be estimated implying a variety of persistent behavior.
The results show very high degree of persistent behavior
till 11~years ($H \approx 1.1$), a slightly less persistent behavior till
100~years ($H \approx 0.8$), and comparatively less persistent behavior
till 2300~years ($H \approx 0.77$).

\begin{acks}
 VS thanks CSIR, India for a Junior Research Fellowship and  AP
acknowledges DST, India for financial support.
\end{acks}

\end{article}

\end{document}